\documentclass[12pt]{article}
\usepackage{ifpdf}
\usepackage{comment}
\usepackage{latexsym}
\usepackage{subfigure}
\usepackage{amssymb}
\usepackage{epsf}
\usepackage{cite}
\usepackage{amsmath}
\ifpdf
\usepackage{hyperref}
\else
\usepackage[hypertex]{hyperref}
\fi
\usepackage{graphicx}% Include figure files
\usepackage{setspace}
\usepackage{subfigure}
\usepackage{slashed}
\usepackage{cancel}
\usepackage{blkarray}

\makeatletter
\let\BA@quicktrue\BA@quickfalse
\makeatother

\makeatletter
    \renewcommand{\theequation}{%
    \thesection.\arabic{equation}}
    \@addtoreset{equation}{section}
  \makeatother

\newcommand{\tr}{{\rm Tr}}
\begin{document}
\pagestyle{empty}

\begin{flushright}
%XXX
\end{flushright}
    \vspace{1cm}
\begin{center}
{\bf\LARGE
Deconstruction, Holography and \\[1ex]
Emergent Supersymmetry
}\\
    \vspace*{2.3cm}
{\bf Yuichiro Nakai} \\
    \vspace*{0.5cm}
{\it Department of Physics, Harvard University, Cambridge, MA 02138}\\
\end{center}
    \vspace*{1.7cm}

\begin{spacing}{1.2}
\begin{abstract}
\vspace{0.2cm}
{\normalsize
We study a gauge theory in a 5D warped space via the dimensional deconstruction
that a higher dimensional gauge theory is constructed from a moose of 4D gauge groups.
By the AdS/CFT correspondence, a 5D warped gauge theory is dual to a 4D conformal field theory (CFT) with a global symmetry.
As far as physics of the gauge theory, we obtain the one-to-one correspondence between each component of a moose of gauge groups
and that of a CFT.
We formulate a supersymmetric extension of deconstruction
and explore the framework of natural supersymmetry in a 5D warped space
-- the supersymmetric Randall-Sundrum model with the IR-brane localized Higgs and bulk fermions -- via the gauge moose.
In this model, a supersymmetry breaking source is located at the end of the moose corresponding to the UV brane and the first two generations of squarks are decoupled.
With left-right gauge symmetries in the bulk of the moose,
we demonstrate realization of accidental or emergent supersymmetry of the Higgs sector in comparison with the proposed ``Moose/CFT correspondence."
}
\end{abstract} 
\end{spacing}

%%%%%%%%%%%%%%%%%%%%%%%%%%%%%%%%%%%%%%%%%%%%%%%%%%%%%%%%%%%%%%%%%%%%%%%%%%%%
\newpage
\baselineskip=17pt
\setcounter{page}{2}
\pagestyle{plain}
\baselineskip=17pt
\pagestyle{plain}

\setcounter{footnote}{0}

\section{Introduction} \label{sec:intro}

The Randall-Sundrum (RS) model
\cite{Randall:1999ee}
of a warped extra dimension has received a lot of attention for many years
from both phenomenological and theoretical perspectives. 
For phenomenology, the RS setup provides one of the leading candidates of physics beyond the Standard Model (SM)
which can address the hierarchy problem.
Although the original model has several problems, its extensions or combinations with other ideas,
such as supersymmetry (SUSY), are still promising.
Theoretically, it is of particular importance that the RS background gives insights into holography.
According to the AdS/CFT correspondence
\cite{AdS/CFT,AdS/CFT2},
the RS warped space with the IR boundary is equivalent to a 4D conformal field theory (CFT)
where the conformal invariance is spontaneously broken at the IR scale
\cite{holography}
(For reviews, see \cite{Aharony:1999ti,Csaki:2005vy,Gherghetta:2010cj}).
This correspondence has been applied to various fields of research, such as studies of QCD
\cite{Son:2003et,QCD}
or models of electroweak symmetry breaking
\cite{electroweak}.

Deconstruction
\cite{deconstruction}
is a useful technique to understand higher dimensional effects of a 5D gauge theory from a purely 4D construction.
In a deconstructed theory, adjacent gauge groups are connected by bi-fundamental scalar fields with nonzero expectation values,
and this theory reproduces physics of the higher dimensional gauge theory at low energies
(See the moose diagram in Figure~\ref{fig:moose} for instance).
In addition, while higher dimensional theories are non-renormalizable and finally rely on the existence of an appropriate UV completion,
deconstruction provides a regularization of them and is a UV completion.
Deconstruction of a gauge theory in a 5D warped space was presented in \cite{Cheng:2001nh,Abe:2002rj,Falkowski:2002cm} (See also \cite{deBlas:2006fz,Burdman:2012sb}).
In the warped gauge theory, running of the gauge coupling is known to be logarithmic
\cite{Pomarol:2000hp}.
Ref.~\cite{Falkowski:2002cm} showed that this behavior is easily understood via deconstruction.

\begin{figure}[!t]
  \begin{center}
  \vspace{0cm}
          \includegraphics[clip, width=10cm]{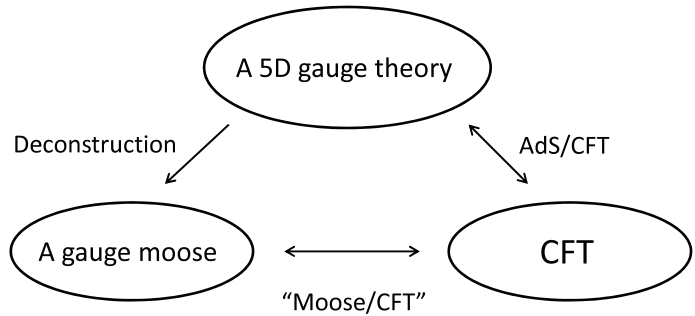}
    \caption{A schematic chart of relations between a gauge theory in a 5D warped space, the deconstructed theory and the dual CFT.}
    \label{fig:Relations}
  \end{center}
  \vspace{0cm}
\end{figure}

We now have two different 4D pictures of a 5D gauge theory -- a moose of gauge groups and a CFT with a global symmetry\footnote{
The symmetry is weakly gauged or not, depending on the UV brane boundary condition in the 5D theory.}
-- in hand.
Figure~\ref{fig:Relations} shows a schematic chart of their relations.
A natural guess is that each component of the gauge moose and that of the CFT are in one-to-one correspondence.
In other words, as far as physics of the gauge theory, it seems to be possible to extract some dictionary of this ``Moose/CFT correspondence'' through the 5D theory.\footnote{
Since large-$N_c$ theories have an infinite number of bound states, the correspondence is only approximate with a finite number of sites in the moose theory.}
In fact, this kind of relation was implied in the study of (large-$N_c$) QCD
\cite{Son:2003et}, motivated by a phenomenological model of the hidden local symmetry
\cite{HLS}.
QCD shows chiral symmetry breaking, spontaneous breaking of the global symmetry in the theory, at low energies.
Holography tells us that a global symmetry in a 4D theory corresponds to a 5D bulk gauge symmetry broken on the UV brane.
By dimensional deconstruction, the 5D gauge theory is described by a moose of gauge groups.
These gauge groups are similar to those which appear in the hidden local symmetries where the gauge bosons correspond to the vector mesons in QCD.
The bi-fundamental scalars connecting the gauge groups form the pions.
The connection was also implied in the discussion of the relation between the Migdal approach
\cite{Migdal:1977nu}
to large-$N_c$ theories and deconstruction
\cite{Falkowski:2006uy} (See also \cite{Shifman:2005zn}).

In this paper, we clarify the statements of the correspondence between a gauge moose and a CFT.\footnote{
Although experts may know most of the statements of the correspondence, the author could not find any literature which describes them explicitly.}
In particular, the relation of the gauge coupling of the moose with the beta function contribution from the CFT is proposed.
We formulate a supersymmetric extension of the deconstruction of a warped gauge theory\footnote{
The deconstruction of a supersymmetric gauge theory in a 5D flat space
and its application to model-building were proposed in Ref.~\cite{Csaki:2001em}
(For other applications, see e.g. \cite{SUSYmodel}). }
and focus on the renormalization group (RG) flow.
One interesting aspect of the gauge theory coupled to a superconformal field theory (SCFT) -- the dual picture of the 5D gauge theory --
is emergence of supersymmetry at low energies
even if SUSY is badly broken in the UV region, as shown in
Ref.~\cite{Sundrum:2009gv}.
We demonstrate it in the gauge moose theory, in comparison with the proposed correspondence between the 4D theories.
Furthermore, the deconstructed version of a realistic supersymmetric RS model with the IR-brane localized Higgs and bulk fermions
is presented.\footnote{
We can take either of the two possibilities that the gauge moose model is realistic in itself
or a useful toy model to describe physics of the warped/composite model.}
Emergent supersymmetry realizes the effective supersymmetry scenario
\cite{effectiveSUSY}
-- consisting of light stops, Higgsinos and gauginos -- with natural electroweak symmetry breaking
when the gaugino masses are protected from large SUSY breaking in some way such as by an approximate R-symmetry.

The rest of the paper is organized as follows.
In section~\ref{sec:deconstruction}, we review the deconstruction of a gauge theory in a 5D warped space.
In section~\ref{sec:Moose/CFT},
the correspondence between a moose of gauge groups and a CFT is proposed.
In section~\ref{sec:warpednaturalSUSY}, we formulate the deconstruction of a supersymmetric RS model and
explore a realistic possibility.
We clarify the problem of a large correction to the Higgs potential through the hypercharge $D$-term
and present the deconstructed version of a possible solution based on a left-right symmetric group
\cite{Sundrum:2009gv,Gherghetta:2011wc,Heidenreich:2014jpa}.
We also comment on Higgs physics in the deconstructed model.
In section~\ref{sec:accidentalSUSY}, we consider the RG flow of the gauge, gaugino and $D$-term couplings in a supersymmetric gauge moose theory
and show realization of emergent supersymmetry.
The degree of fine-tuning is estimated in the realistic model presented in the former section.
In section~\ref{sec:conclusions}, we conclude discussions and comment on future directions.
In the appendix, the deconstruction of bulk matter fields is summarized.

\section{Deconstruction of a warped gauge theory} \label{sec:deconstruction}

We concentrate on the following background geometry in this paper: a 5D warped space whose extra dimension is compactified on an interval,
$0 \leq y \leq \pi R$.
The UV (IR) brane is located at the end of the interval, $y = 0$ ($y = \pi R$). 
The spacetime metric of a slice of $\rm AdS_5$ is presented as~\cite{Randall:1999ee}
\begin{equation}\label{metric}
\begin{split}
\\[-2.5ex]
ds^2 \, = \, e^{-2k y} \eta_{\mu\nu} dx_\mu dx_\nu + dy^2 \, ,\\[1.5ex]
\end{split}  
\end{equation}
where $\mu, \nu = 0, \cdots, 3$ are the usual 4D indices, $\eta_{\mu\nu} = {\rm diag} (-1,1,1,1)$ and $k$ is the AdS curvature.
We assume $e^{- \pi k R} \ll 1$ as in the original Randall-Sundrum model.

We consider a $U(1)$ gauge theory on the above background and latticize it.
Generalization to non-Abelian gauge theories is straightforward.
The continuum action of the 5D gauge field $A_M = (A_\mu, A_5)$ is given by
\begin{equation}\label{continuum}
\begin{split}
\\[-2.5ex]
S_5 \,=\, \int d^4 x \int dy \, \frac{1}{g_5^2} \left( - \frac{1}{4} F_{\mu\nu} F_{\mu\nu} - \frac{e^{-2ky}}{2} \left( \partial_\mu A_5 - \partial_5 A_\mu \right)^2 \right), \\[1.5ex]
\end{split}
\end{equation}
where $g_5$ is the 5D gauge coupling with mass dimension $-1/2$.
In the rest of the discussion, we concentrate on the following boundary condition of the gauge field on the UV brane:
\begin{equation}
\begin{split}
\\[-2.5ex]
\partial_5 A_\mu (0) \,=\, A_5 (0) \,=\, 0 \, , \\[1.5ex]
\end{split}
\end{equation}
that is, $A_\mu$ ($A_5$) satisfies the Neumann (Dirichlet) condition on the UV brane.
We divide the extra dimension into $N$ intervals with the lattice spacing $a$ so that the coordinate $y$ is composed of lattice points $y_j$ $\left( j = 0, \cdots, N \right)$
where $y_{0, \, N}$ correspond to $y=0, \pi R$ respectively ($Na = \pi R$).
The action \eqref{continuum} is then reduced to
\begin{equation}\label{latticized}
\begin{split}
\\[-2.5ex]
S_5 \,\rightarrow\, \int d^4 x \, &\frac{a}{g_5^2} \, \biggl( - \frac{1}{4} \sum_{j=0}^{N} F_{\mu\nu} (y_j) F_{\mu\nu} (y_j) \\[1ex]
&- \frac{1}{2} \sum_{j=1}^{N} e^{-2 k y_j} \left( \partial_\mu A_5 (y_j) - \frac{A_\mu (y_j) - A_\mu (y_{j-1})}{a} \right)^2 \, \biggr) \, . \\[1ex]
\end{split}
\end{equation}
Note that this action is invariant under the transformations,
\begin{equation}
\begin{split}
\\[-2.5ex]
y_j \, &\rightarrow \, y_j + a \, , \\[1.5ex]
x_\mu \, &\rightarrow \, e^{ka} x_\mu \, , \\[1.5ex]
\partial_\mu \, &\rightarrow \, e^{-ka} \partial_\mu \, , \\[1.5ex]
A_\mu \, &\rightarrow \, e^{-ka} A_\mu \, , \label{conformalsim} \\[1.5ex]
\end{split}
\end{equation}
for $j= 0, \cdots, N-1$, reflecting the fact that the $\rm AdS_5$ space has the 4D conformal group $SO(2, 4)$ as the isometry group.
The lattice point $y_N$ breaks this symmetry as the IR brane does.

\begin{figure}[!t]
  \begin{center}
  \vspace{-0.5cm}
          \includegraphics[clip, width=16cm]{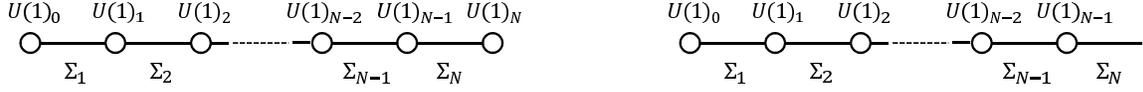}
          \vspace{-0.2cm}
    \caption{The moose diagram of a $U(1)^{N+1}$ ($U(1)^{N}$) gauge theory with $N$ complex scalars
when the corresponding 5D gauge field $A_\mu$ satisfies the Neumann (Dirichlet) condition on the IR brane.}
    \label{fig:moose}
  \end{center}
\end{figure}

We now present the deconstructed model of the above 5D gauge theory
\cite{Cheng:2001nh,Abe:2002rj,deBlas:2006fz,Falkowski:2002cm}.
When the 5D gauge field $A_\mu$ ($A_5$) satisfies the Neumann (Dirichlet) condition on the IR brane,
the model is given by a $U(1)^{N+1}$ gauge theory with $N$ complex scalars called the link fields $\Sigma_j$
where the index $j$ runs from $1$ to $N$.
Each link field $\Sigma_j$ has charges $(- 1, + 1)$ under the $U(1)_{j-1} \times U(1)_j$ subgroup.
The moose diagram is shown on the left of Figure~\ref{fig:moose}.
The action is denoted as
\begin{equation}\label{deconstructed action}
\begin{split}
\\[-2ex]
S_4 \, = \, \int d^4 x \,  \biggl( - \frac{1}{4} \sum_{j=0}^{N} \frac{1}{g_j^2} F^j_{\mu\nu} F^j_{\mu\nu}
- \sum_{j=1}^{N} \left| \partial_\mu \Sigma_j + i \left( A_\mu^{j-1} - A_\mu^j \right) \Sigma_j \right|^2 - V ( |\Sigma_j|^2 ) \biggr) \, , \\[1.5ex]
\end{split}
\end{equation}
where $g_j$ is the gauge coupling of the group $U(1)_j$.
The potential $V ( |\Sigma_j|^2 )$ stabilizes the link fields at $\langle \Sigma_j \rangle = v_j / \sqrt{2}$.
To reproduce the gauge theory on the RS warped background \eqref{metric},
the vacuum expectation value (VEV) of the link field $v_j$ and the gauge coupling $g_j$ defined at the scale $v_j$ are chosen as
\begin{equation}\label{vev}
\begin{split}
\\[-2ex]
v_j = v e^{-kj/gv} \, , \qquad g_j = g_j (v_j) = g \, . \\[1.5ex]
\end{split}
\end{equation}
We also assume $g_0 \, (v) = g $ here for simplicity, but we will generalize it in the later section.
Due to the nonzero VEVs of the scalars, the $U(1)^{N+1}$ gauge symmetry is broken down to a $U(1)$ at low energies
and the Higgs mechanism gives masses to the gauge bosons corresponding to the broken generators.
One massless component of each $\Sigma_j$ is the would-be Nambu-Goldstone mode eaten by the massive gauge field.
We ignore the massive component and take a parametrization $\Sigma_j \rightarrow v_j e^{i \xi_j/ v_j}$.
The action of the 4D theory is then rewritten as
\begin{equation}\label{deconstructed action2}
\begin{split}
\\[-2.5ex]
S_4 \, \rightarrow \, \int d^4 x \,  \biggl( - \frac{1}{4g^2} \sum_{j=0}^{N} F^j_{\mu\nu} F^j_{\mu\nu}
- \frac{1}{2} \sum_{j=1}^{N}  \left[ \partial_\mu \xi_j - v_j \left( A_\mu^{j} - A_\mu^{j-1} \right)  \right]^2 \biggr) \, . 
\end{split}
\end{equation}

Let us compare the latticized 5D action \eqref{latticized} with the deconstructed action \eqref{deconstructed action2}.
The dictionary of the correspondence between the 5D gauge theory and the deconstructed theory is then given  by
\begin{equation}
\begin{split}
N (\pi R)^{-1} = a^{-1} \, &\leftrightarrow \, gv \,,  \\[1ex]
y_j \, a^{-1} \, &\leftrightarrow \, j \,,  \\[1ex]
g_5 \, a^{-1/2} \, &\leftrightarrow \, g \,,  \\[1ex]
A_\mu (y_j) \, &\leftrightarrow \, A_\mu^j \,,  \\[1ex]
e^{-ky_j} A_5 (y_j) \, &\leftrightarrow \, g \xi_j \,. \label{dictionary} \\[2ex]
\end{split}
\end{equation}
In the present work, we consider a coarse lattice, assuming $a^{-1} \sim k$ (or equivalently $N \sim \pi k R$)
so that $e^{-k/gv} \leftrightarrow e^{-k a} \ll 1$.
Then, from \eqref{vev}, we have
\begin{equation}\label{vev2}
\begin{split}
\\[-2.5ex]
v_1 \gg v_2 \gg \cdots \gg v_{N-1} \gg v_N \, , \\[1.5ex]
\end{split}
\end{equation}
which will greatly simplify our analyses.
In the case of a fine lattice, $a^{-1} \gg k$,
the VEVs of adjacent link fields are no longer hierarchical and the corresponding extra dimension is locally flat
(but globally warped).

We now analyze the mass spectrum of the gauge bosons in the deconstructed model.
Canonically normalizing the kinetic terms, the action \eqref{deconstructed action2} contains the gauge boson mass terms,
\begin{equation}\label{massterm}
\begin{split}
\mathcal{L}_4 \, \supset \, - \frac{1}{2} \sum_{j=1}^{N} \,  ( g v_j )^2 \left( A_\mu^{j} - A_\mu^{j-1} \right)^2 . \\[1ex]
\end{split}
\end{equation}
Although it is difficult to diagonalize the full mass matrix precisely,
due to the hierarchical VEVs \eqref{vev2}, we can obtain the approximate mass spectrum
by diagonalizing the mass matrix of two adjacent $U(1)$ subgroups step by step.
First, we consider the gauge fields $A_\mu^0$ and $A_\mu^1$.
The VEV $v_1$ leads to the gauge symmetry breaking $U(1)_0 \times U(1)_1 \rightarrow U(1)_{(1)}$
(We denote the diagonal subgroup as $U(1)_{(\cdot)}$).
One eigenstate has a mass of order, $m_{v_1} \sim g v_1$, while the other remains massless.
Next, at the scale $v_2$, the breaking $U(1)_{(1)} \times U(1)_2 \rightarrow U(1)_{(2)}$ occurs and a gauge boson becomes massive,
$m_{v_2} \sim g v_2$.
The breaking occurs repeatedly at each scale of the link field VEVs.
When the 5D gauge field $A_\mu$ ($A_5$) satisfies the Neumann (Dirichlet) condition on the IR brane,
we finally obtain one massless eigenstate,
\begin{equation}
\begin{split}
A_\mu^{(N)} =  \frac{1}{\sqrt{N+1}} \sum_{j=0}^N A_\mu^j \, ,  \\[1.5ex]
\end{split}
\end{equation}
which corresponds to the zero mode in the Kaluza-Klein (KK) decomposition of the 5D gauge field.

When the 5D gauge field $A_\mu$ ($A_5$) satisfies the Dirichlet (Neumann) condition on the IR brane,
all we have to do is to remove the $U(1)_N$ subgroup at the end of the moose
(See the diagram on the right of Figure~\ref{fig:moose}).
The action is given as with \eqref{deconstructed action} except $A_\mu^N = 0$.
In this case, the nonzero VEVs of the link fields break the $U(1)^{N}$ gauge symmetry completely
and all the gauge bosons become massive.

\section{The Moose/CFT correspondence} \label{sec:Moose/CFT}

The AdS/CFT correspondence
\cite{AdS/CFT,AdS/CFT2,holography} tells us that
the bulk of AdS is equivalent to a CFT
(See \cite{Aharony:1999ti,Csaki:2005vy,Gherghetta:2010cj} for reviews).
A bulk gauge symmetry is understood as a (weakly gauged) global symmetry in the CFT.
As we have seen in the last section,
the warped gauge theory can be also described in terms of a moose of gauge groups.
It is naturally expected that
there is a relation between the CFT and the gauge moose and a gauge symmetry at each site of the moose corresponds to a (weakly gauged) global symmetry in the CFT.
We here clarify this correspondence further through the 5D theory.
In particular, the proposed relation between the gauge coupling of the moose and the beta function contribution from the CFT
is useful for the discussion in section~\ref{sec:accidentalSUSY}.

First, let us consider the symmetry respected in both 4D pictures.
As noted in the previous section,
we have a one-to-one correspondence between a latticized 5D gauge theory and the deconstructed theory.
The latticized 5D theory respects a discretized subset of the conformal group $SO(2,4)$ as in \eqref{conformalsim}.
Since the large-$N_{\rm CFT}$ theory dual to the 5D theory respects the continuum conformal symmetry,
the correspondence between the deconstructed theory and the CFT is only approximate
(In the fine lattice limit, the correspondence is exact as far as physics of the gauge theory).
Next, we focus on two ends of the moose.
We call the end $j=0$ ($j=N$) corresponding to the UV (IR) brane as the UV (IR) end.
Since appearance of the UV brane is understood as a UV cutoff of the CFT,
the mass scale $g v \sim k$ in the moose theory corresponds to a cutoff scale of the CFT (denoted as $m_0$).
Matter fields only coupled to the gauge group at the UV end
are understood as elementary fields coupled to the CFT.
On the other hand,
appearance of the IR brane spontaneously breaks the conformal symmetry,
and hence the breaking scale $\Lambda_{\rm comp}$ corresponds to the lightest massive gauge boson mass scale $g v_N$.
Matter fields localized at the IR end are composite fields in the CFT spectrum.
The spectrum of the moose theory approximately describes the spectrum of the CFT. 

Let us rephrase the above correspondence in a slightly more abstract form.
According to the AdS/CFT correspondence, there exists an associated CFT operator for every 5D bulk field
and the boundary value of the bulk field is considered as a source field for the CFT operator.
The correspondence is quantified by the famous generating functional
\cite{AdS/CFT2}.
In the ``Moose/CFT correspondence," we have a similar relation such as
\begin{equation}
\begin{split}
\\[-2ex]
Z[\phi^0] \, = \, \int \mathcal{D} \phi_{\rm CFT} \, e^{i S_{\rm CFT} [\phi_{\rm CFT}] + i \int d^4 x \, \phi^0 \mathcal{O}} \, = \,
\int_{\phi^0} \Pi_j \, \mathcal{D} \phi^j \, e^{i S [\phi^j] } \, \equiv \,  e^{iS_{\rm eff} [\phi^0] } \, , \label{witten} \\[1ex]
\end{split}
\end{equation}
where $S_{\rm CFT} [\phi_{\rm CFT}]$ is the action of the CFT fields $\phi_{\rm CFT}$,
$\mathcal{O}$ is the CFT operator associated with a moose of fields $\phi^j$ and
$\phi^0$ is the field at the UV end of the moose.\footnote{
In our case, $\phi^0$ is also a dynamical field.}
Here, $\phi^j$ can be a gauge field or a matter field put at every site of a gauge moose (See the appendix).
A source term $\phi^0 \mathcal{O}$ is added to the CFT action.
$S [\phi^j]$ is the action of the moose theory whose boundary value is fixed by the source $\phi^0$.
Integrating out $\phi^j$ degrees of freedom except for $\phi^0$ by using their equations of motion,
we can obtain the effective action $S_{\rm eff}$ which is the functional of $\phi^0$.
The $n$-point function of the CFT operator $\mathcal{O}$ is then given by
\begin{equation}
\begin{split}
\langle \mathcal{O} \dots \mathcal{O} \rangle \, = \, \frac{\delta^{n} S_{\rm eff} }{\delta \phi^0 \dots \delta \phi^0} \, . \\[1ex]
\end{split}
\end{equation}
The $n$-point functions of the strongly coupled theory can be calculated from a moose theory.

To illustrate the correspondence with an example,
we take a $U(1)^{N+1}$ gauge theory presented in the previous section (See Figure~\ref{fig:moose})
and calculate the effective action -- the functional of $A_\mu^0$ --
to give the $n$-point functions of the corresponding CFT operator
\cite{Falkowski:2006uy}.
Moving to the momentum space, the action of the moose theory \eqref{deconstructed action2} is given by
\begin{equation}
\begin{split}
\\[-2.5ex]
S_4 \, \rightarrow \, \frac{1}{2} \int \frac{d^4 p}{(2\pi)^4} \left( \, \sum_{j, \, k = 0}^N A_\mu^j D_{\mu\nu}^{jk} A_\nu^k \right) ,  \label{momentumaction} \\[1ex]
\end{split}
\end{equation}
where the kinetic operator is defined as
\begin{equation}
\begin{split}
\\[-2.5ex]
D_{\mu\nu}^{jk} \, = \,  \left( -p^2 \eta_{\mu\nu} + p_\mu p_\nu \right) \frac{1}{g^2} \, \delta_{j, \, k} - \eta_{\mu\nu} \left( (v_j^2 + v_{j+1}^2) \, \delta_{j, \, k}
- v_j^2 \, \delta_{j-1, \, k} - v_{j+1}^2 \delta_{j+1, \, k} \right) , \\[1ex]
\end{split}
\end{equation}
with $v_0, \, v_{N+1} = 0$ (We here take $\xi_j = 0$).
To obtain the effective action $S_{\rm eff}$,
we now integrate out $A_\mu^j$ degrees of freedom except for $A_\mu^0$
by using their equations of motion,
\begin{equation}
\begin{split}
\sum_{k=0}^N D_{\mu\nu}^{jk} A_{\nu}^k \, = \, 0 \, , \qquad j \geq 1 \, . \\[0ex]
\end{split}
\end{equation}
Solving these equations of motion and inserting the solutions into the original action \eqref{momentumaction},
we can obtain the effective action as the functional of $A_\mu^0$,
\begin{equation}
\begin{split}
\\[-2.5ex]
S_{\rm eff} \, = \, \int \frac{d^4 p}{(2\pi)^4} \, v_1^2 \, A_\mu^0 (p) \left( \eta_{\mu\nu} -\frac{p_\mu p_\nu}{p^2} \right) A_\nu^0 (p) \, \Pi(p^2) \, . \\[1ex]
\end{split}
\end{equation}
The polarization operator is given by
\begin{equation}
\begin{split}
\\[-2.5ex]
\Pi (p^2) \, = \, \frac{1}{F^0_N (p^2)} \left\{  F^1_N (p^2) - F^0_N (p^2) \left(  \frac{1}{g^2 v_1^2} \, p^2 + 1 \right) \right\} \, , \label{polarization} \\[1ex]
\end{split}
\end{equation}
where $F^j_N$ is a solution of the equation,
\begin{equation}
\begin{split}
\\[-2.5ex]
\left( v_{j+1}^2 + v_j^2 + \frac{p^2}{g^2} \right) F_N^j - v_j^2 F^{j-1}_N - v_{j+1}^2 F_N^{j+1} \, = \, 0 \, , \qquad j \geq 1 \, . \\[1ex]
\end{split}
\end{equation}
With a constant $F^N_N$, the solution $F_N^j (p^2)$ is a polynomial of degree $N-j$.
We can calculate the $n$-point functions of the CFT operator from the obtained effective action.
In particular, the mass spectrum of the CFT states is given by zeros of the polarization operator \eqref{polarization}.
Due to the hierarchical VEVs \eqref{vev2}, this is roughly given by
\begin{equation}
\begin{split}
\\[-2.5ex]
\Bigl\{ \, \dots  \Bigr\} \, \sim \, - \frac{p^2}{g^2 v_1^2} \left( 1 + \frac{p^2}{g^2 v_1^2} \right)
\left( 1 + \frac{p^2}{g^2 v_2^2} \right) \cdots \left( 1 + \frac{p^2}{g^2 v_N^2} \right) , \\[1ex]
\end{split}
\end{equation}
which leads to just the mass spectrum of the gauge bosons obtained in the previous section.

Finally, we present the correspondence between the gauge coupling of a gauge moose theory and the beta function coefficient contribution from the dual CFT.
In the 5D theory, tree-level matching of the 5D gauge coupling $g_5$ with the effective 4D gauge coupling $g_4$ leads to
\begin{equation}
\begin{split}
\frac{1}{g_4^2} \, &= \,  \frac{\pi R}{g_5^2} \\[1ex]
\, &= \,  \frac{1}{g_5^2 k} \ln \left( \frac{m_0}{\Lambda_{\rm comp}} \right) , \\[1.5ex]
\end{split}
\end{equation}
where $m_0 = \Lambda_{\rm comp} \, e^{\pi k R}$.
This represents that the coordinate along the AdS is the energy scale in the CFT
and the inverse of the (dimensionless) gauge coupling squared $g_5^2 k$ corresponds to 
the beta function contribution from the CFT sector, $b_{\rm CFT} \leftrightarrow 1/(g_5^2 k)$.
Then, using the dictionary \eqref{dictionary}, the correspondence between $b_{\rm CFT}$ and the gauge coupling of a gauge moose theory $g$ is given by
\begin{equation}
\begin{split}
\\[-2.5ex]
b_{\rm CFT} \,\leftrightarrow\, \frac{1}{g^2} \left( \frac{k}{gv} \right)^{-1} . \label{bcft/gauge} \\[1.5ex]
\end{split}
\end{equation}
Since we have assumed $gv \sim k$, the inverse of the gauge coupling squared of the moose theory is roughly the beta function contribution, $b_{\rm CFT} \sim 1/g^2$.
The coefficient $b_{\rm CFT}$ is of order $N_{\rm CFT}/(16\pi^2)$ and hence $g^2 = \mathcal{O} (16\pi^2 / N_{\rm CFT})$.
The relation \eqref{bcft/gauge} has an important role in comparing the RG flow of a moose theory with that of a gauge theory coupled to a CFT,
as discussed in section~\ref{sec:accidentalSUSY}.

\section{Deconstruction of SUSY RS} \label{sec:warpednaturalSUSY}

In this section, we present the deconstruction of supersymmetric gauge theories in a 5D warped space and
apply it to a realistic framework with the IR-brane localized Higgs and bulk fermions, which we call warped natural SUSY
\cite{Sundrum:2009gv,Gherghetta:2011wc,Heidenreich:2014jpa}
(See also \cite{Gherghetta:2003he,Larsen:2012rq}).
In this setup, there is a potentially large correction to the Higgs potential through the hypercharge $D$-term
\cite{Sundrum:2009gv,Strassler:2003ht}.
We clarify this $U(1)$ $D$-term problem in terms of the deconstructed theory and present a solution by a left-right symmetric gauge moose.
We also comment on Higgs physics in deconstructed warped natural SUSY.

\subsection{Supersymmetrization} \label{subsec:SUSYRS}

\begin{figure}[!t]
  \begin{center}
  \vspace{-0.5cm}
          \includegraphics[clip, width=16cm]{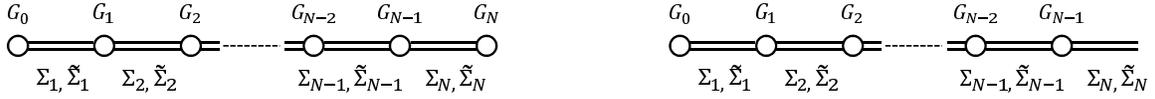}
          \vspace{-0.2cm}
    \caption{The moose diagram of a supersymmetric $G^{N+1}$ ($G^{N}$) gauge theory with $N$ vector-like pairs of link chiral superfields $\Sigma_j$, $\widetilde{\Sigma}_j$
when the corresponding 5D gauge field $A_\mu$ satisfies the Neumann (Dirichlet) condition on the IR brane.}
    \label{fig:SUSYMoose}
  \end{center}
\end{figure}

Let us consider a supersymmetric extension of the deconstructed theory presented in section~\ref{sec:deconstruction}.
We here denote the gauge group as $G$ generally.
When the 5D gauge field $A_\mu$ ($A_5$) satisfies the Neumann (Dirichlet) condition on the IR brane,
the model is given by a supersymmetric $G^{N+1}$ gauge theory with $N$ vector-like pairs of link chiral superfields $\Sigma_j$, $\widetilde{\Sigma}_j$
(The reason of introducing the conjugate chiral superfields $\widetilde{\Sigma}_j$ is explained below\footnote{
Ref.~\cite{Abe:2002rj} discusses a supersymmetric extension of deconstructed warped gauge theories without introducing the conjugate chiral superfields.}).
The gauge field $A_\mu^j$ at each site of the moose now becomes a component of an $\mathcal{N} =1$ vector multiplet $V^j$.
Each link field $\Sigma_j$ has the anti-fundamental and fundamental representations under the $G_{j-1}$ and $G_j$ subgroups respectively.
The moose diagram is shown on the left of Figure~\ref{fig:SUSYMoose}.
The supersymmetric Lagrangian is denoted as
\begin{equation}
\begin{split}
\\[-2.5ex]
\mathcal{L}_{4, \, \rm SUSY} \, &= \, \frac{1}{4} \int d^2 \theta \, \sum_{j=0}^N \,  \mathcal{W}_\alpha^{\,j} \mathcal{W}_\alpha^{\,j} \, + \, {\rm h.c.} \\[1ex]
&\quad+ \int d^4 \theta \, \sum_{j=1}^N \left(  \, {\rm Tr} \, e^{ - g_j V^j} \Sigma_j \, e^{ \, g_{j-1} V^{j-1}} \Sigma_j^\dagger
+ {\rm Tr} \, e^{-g_{j-1} V^{j-1}} \widetilde{\Sigma}_j \, e^{ \, g_j V^j} \widetilde{\Sigma}_j^\dagger \, \right)   \\[1ex]
&\quad+ \int d^2 \theta \, W(\Sigma_j, \widetilde{\Sigma}_j) \, , \\[1ex]
\end{split}
\end{equation}
where the superpotential $W(\Sigma_j, \widetilde{\Sigma}_j)$ stabilizes the link scalar fields at $\langle \Sigma_j \rangle = \langle \widetilde{\Sigma}_j \rangle = \frac{v_j}{\sqrt{2}} {\bf 1}$.
As in the non-supersymmetric case,
the gauge coupling $g_j$ defined at the scale $v_j$ is chosen as $g_j = g_j (v_j) = g$ to reproduce the gauge theory on the RS warped background \eqref{metric}.
The $G^{N+1}$ gauge symmetry is broken down to a $G$ at low energies
and the gauge fields corresponding to the broken generators form massive vector multiplets, absorbing the link chiral superfields.
The dictionary of the correspondence between the (latticized) 5D gauge theory and the deconstructed theory is the same as that given in \eqref{dictionary}.
The Moose/CFT relations discussed in the previous section are also extended to accommodate supersymmetry without any difficulty.
We consider a coarse lattice so that $e^{-k/gv} \leftrightarrow e^{-k a} \ll 1$ as well.
There are two reasons to introduce the conjugate chiral superfields $\widetilde{\Sigma}_j$.
Since the link chiral multiplets contain fermion components, it is required to make the theory anomaly-free.
In addition, due to the hierarchical VEVs \eqref{vev2}, the $D$-term potential of the link fields $\Sigma_j$ does not vanish without the conjugate fields $\widetilde{\Sigma}_j$
(This problem does not occur in the case of the flat extra dimension where all the VEVs of the link fields are equal).
As in the non-supersymmetric case, when the 5D gauge field $A_\mu$ ($A_5$) satisfies the Dirichlet (Neumann) condition on the IR brane,
the $G_N$ subgroup at the end of the moose is removed from the model
(See the diagram on the right of Figure~\ref{fig:SUSYMoose}).

A supersymmetric extension of the deconstruction of bulk matter fields is briefly presented in the appendix.
A 5D bulk matter multiplet -- an $\mathcal{N} = 2$ hypermultiplet in 4D terms -- is described
in terms of a vector-like pair of chiral superfields at every site except for the UV end.
At this end site, only a chiral superfield is put to give a chiral zero mode multiplet.
According to the correspondence described in the last section, this chiral superfield provides a source supermultiplet for the associated CFT operator.

\subsection{Deconstructed warped natural SUSY} \label{subsec:warpednatural}

\begin{figure}[!t]
  \begin{center}
  \vspace{-1cm}
          \includegraphics[clip, width=7cm]{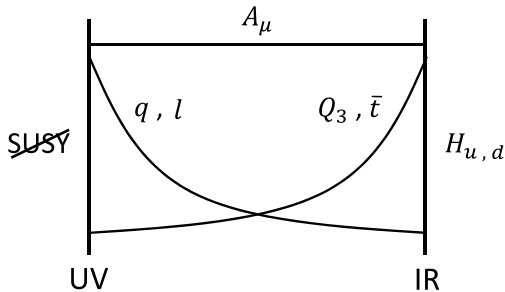}
  \vspace{0.3cm}
    \caption{A schematic picture of warped natural SUSY.
The Higgs doublets $H_{u, \, d}$ and their superpartners live on the IR brane.
The SM gauge bosons (denoted as $A_\mu$) and the gauginos propagate in the bulk of the extra dimension.
Light quarks $q$, leptons $l$ and their superpartners are localized toward the UV brane while heavy (s)quarks $Q_3, \bar{t}$ are localized toward the IR brane
(There are two possibilities for the locations of 
the right-handed bottom quark and tau lepton multiplets: they can be localized toward the UV or toward the IR brane).
A SUSY breaking source is put on the UV brane.}
    \label{fig:Scenario}
  \end{center}
\end{figure}

Combining the paradigms of supersymmetry and the RS model provides a convincing possibility beyond the SM.
In the framework of warped natural SUSY
\cite{Sundrum:2009gv,Gherghetta:2011wc,Heidenreich:2014jpa},
the Higgs fields live on the IR brane while the
SM gauge multiplets and the quark and lepton multiplets propagate in the bulk of the extra dimension,
so that the Yukawa hierarchies arise from their wavefunction profiles
\cite{Yukawa,Gherghetta:2000qt} (We have two reasonable possibilities for the locations of the third generation matter fields:
the right-handed bottom quark and tau lepton multiplets are localized toward the UV or toward the IR brane).
A schematic picture is given in Figure~\ref{fig:Scenario}.
A SUSY breaking source is assumed to be localized on the UV brane.
Since the zero modes of light quark and lepton multiplets live near the UV brane, they can directly couple to the SUSY breaking source
and the squarks and sleptons are heavy enough to avoid excessive flavor and CP violation.
Just below their mass scale, supersymmetry is badly broken.
Nonetheless, SUSY re-emerges near the IR brane
\cite{Gherghetta:2003he} (except for a potential problem from the hypercharge $D$-term, as discussed later).
Since the Higgs multiplets and the top (s)quark zero modes are localized toward the IR brane,
they hardly couple to the SUSY breaking source.
Although SUSY breaking is mediated by the bulk gauge multiplets at 1-loop, the effect on the multiplets localized toward the IR brane is accidentally small, as proven in the next section.
Their soft masses are also generated from other effects below the IR scale such as gaugino mediation
\cite{gauginomed}.
Then, natural electroweak breaking is realized when the gaugino masses are protected from large SUSY breaking
in some way such as by an approximate R-symmetry.

We now describe the deconstructed version of this framework.
Since the 5D theory has the SM gauge supermultiplets propagating in the bulk of the extra dimension,
the gauge group at every site of the moose is given by $G = SU(3)_C \times SU(2)_L \times U(1)_Y$.
$N$ vector-like pairs of link chiral superfields are introduced for each of the $SU(3)_C$, $SU(2)_L$ and $U(1)_Y$ groups.
The charge of the $U(1)_Y$ link field is not uniquely determined
if we do not impose some assumption such that the link field forms some unified multiplet with the $SU(3)_C$ and/or $SU(2)_L$ link fields.
The Higgs fields live on the IR brane in the 5D model and hence they are introduced at the IR end of the moose.
Each multiplet of quarks and leptons in 5D is described
in terms of a vector-like pair of chiral superfields at every site except for the UV end.
As commented above, only a chiral superfield is put to give a chiral zero mode multiplet at this end.
The bulk mass parameters of the matter multiplets are chosen to reproduce the Yukawa hierarchies 
\cite{Abe:2002rj}
(See also the appendix).
A SUSY breaking source is put at the UV end of the gauge moose.

\begin{figure}[!t]
  \begin{center}
  \vspace{-0.5cm}
          \includegraphics[clip, width=10cm]{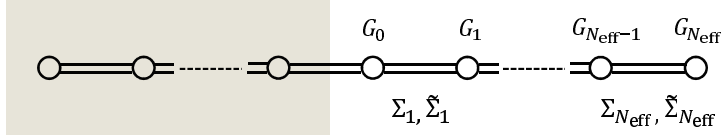}
          \vspace{0.2cm}
    \caption{The moose diagram of a supersymmetric $G^{N_{\rm eff} +1}$ gauge theory with $N_{\rm eff}$ vector-like pairs of link chiral superfields for each factor of $G$
as the effective moose theory below the heavy scalar mass scale.
At the effective UV end $j=0$, light quarks and leptons do not have their scalar superpartners.}
    \label{fig:EffectiveMoose}
  \end{center}
\end{figure}

The moose of gauge groups is broken to a smaller moose with one fewer site at the scale of each VEV.
The gauge fields corresponding to the broken generators become massive and are integrated out below the breaking scale.
The matter multiplets at the broken gauge site also become massive and are integrated out supersymmetrically until the mass scale of heavy squarks and sleptons is reached.
Below this scale, we obtain an effective moose theory,
a supersymmetric $G^{N_{\rm eff} +1}$ gauge theory with $N_{\rm eff}$ vector-like pairs of link chiral superfields for each factor of $G$, whose
effective UV end has light quarks and leptons without their superpartners (See Figure~\ref{fig:EffectiveMoose}).
The theory is apparently non-supersymmetric at this scale.
We now redefine $G_0$ as the gauge group at the effective UV end and 
denote $N_{\rm eff}$ as just $N$ for simplicity of notation in the rest of this paper.
The results in section~\ref{sec:deconstruction}, \ref{sec:Moose/CFT} are applied without change in this redefinition.
The scale $g v \sim k$ in the effective moose theory corresponds to the cutoff scale of the dual CFT $m_0$ as before.
We comment at each time when $N$ indicates the original number of the gauge sites.

In general, there is a fast proton decay problem in RS models due to the low cutoff scale on the IR brane
\cite{Gherghetta:2000qt}.
In the supersymmetric extension, even if we impose R-parity as usual,
we can write a dangerous dimension five operator $\frac{1}{\Lambda_{\rm comp}} QQQL$.
In the 5D setup, this problem is solved by imposing a lepton or baryon number symmetry on the model
(See Ref.~\cite{Heidenreich:2014jpa}).
On the other hand, the deconstructed theory provides a UV completion above the IR scale.
Hence, if we only impose R-parity and assume the cutoff scale of the theory is high enough, e.g. the Planck scale,
the proton decay problem can be avoided with the profiles of quark and lepton zero modes.

\subsection{D-terms and gauge extensions} \label{subsec:D-terms}

In the effective moose theory after integrating out the heavy scalars, a large Fayet-Iliopoulos (FI) term is generated 
at the UV end of the $U(1)_Y$ gauge moose,
\begin{equation}
\begin{split}
\\[-2.5ex]
\mathcal{L}_{\rm FI} \, = \, \int d^4 \theta \,\,  2 \xi \, V^0 \, , \label{FI} \\[1ex]
\end{split}
\end{equation}
where $V^0$ is the vector superfield of the $U(1)_{Y 0}$ gauge group and
$\xi \sim \frac{g_Y}{16\pi^2} m_0^2 \,$.
The $U(1)_Y$ $D$-term potential of the link scalars and the Higgs fields at the IR end is then given by
\begin{equation}
\begin{split}
\\[-2.5ex]
V_D \, &= \, \frac{g_Y^2}{2} \left( \, |\Sigma_1|^2 -  |\widetilde{\Sigma}_1 |^2 - \frac{\xi}{g_Y} \, \right)^2 \\[1ex]
&\quad+ \frac{g_Y^2}{2} \left( \, |\Sigma_1|^2 - |\widetilde{\Sigma}_1|^2 - |\Sigma_2|^2 + |\widetilde{\Sigma}_2|^2 \, \right)^2 \\[1ex]
&\quad+ \cdots \\[1ex]
&\quad+ \frac{g_Y^2}{2} \left( \, |\Sigma_N|^2 - |\widetilde{\Sigma}_N|^2 - \frac{1}{2} |H_u|^2 + \frac{1}{2} |H_d|^2 \right)^2, \label{DtermL} \\[1ex]
\end{split}
\end{equation}
where we have assumed that the $U(1)_Y$ link fields have unit charges.
The potential is stabilized at
\begin{equation}
\begin{split}
\\[-2.5ex]
|\Sigma_j|^2 - |\widetilde{\Sigma}_j|^2 \, = \, \frac{\xi}{g_Y} \, , \qquad |H_u|^2 - |H_d|^2 \, = \, \frac{2\xi}{g_Y} \, , \\[1ex]
\end{split}
\end{equation}
for every $j$, and electroweak symmetry is badly broken because the other mass parameters in the Higgs potential are at the electroweak scale smaller than a large $\sqrt{\xi}$.

To solve this problem, we have to forbid the FI term \eqref{FI} in some way.
One possible solution is to extend the SM gauge groups at each site of the moose to some unified group such as a left-right symmetric group,
as proposed in the 5D model \cite{Sundrum:2009gv,Gherghetta:2011wc,Heidenreich:2014jpa}.
The minimal left-right symmetry
\cite{Mohapatra:1974gc}
is given by $SU(3)_C \times SU(2)_L \times SU(2)_R \times U(1)_{B-L}$ where
the $U(1)_{B-L}$ $D$-term changes sign under the left-right symmetry and hence the FI term is forbidden.
At the IR end, only the SM gauge groups are realized, reflecting the fact that
the unified group is broken on the IR brane by boundary conditions:
the extra gauge fields $A'_\mu$ ($A'_5$) satisfy the Dirichlet (Neumann) boundary condition.
In this case,
all the extra gauge bosons of the extended gauge group become massive.
The lightest modes, $W'$ and $Z'$ for the left-right symmetry, have masses of order,
\begin{equation}
\begin{split}
\\[-2.5ex]
m_{W'} \, \sim \, \sqrt{\frac{1}{N}} \, g v_N \, , \\[1.5ex]
\end{split}
\end{equation}
where $N$ indicates the original number of the sites, instead of $N_{\rm eff}$.
As expected, this mass can be also understood in terms of
radiative corrections from the CFT states (See Ref.~\cite{Heidenreich:2014jpa}),
\begin{equation}
\begin{split}
\\[-2.5ex]
m_{W'} \, \sim \, \sqrt{\frac{1}{\pi k R}} \, \Lambda_{\rm comp} \, , \\[1.5ex]
\end{split}
\end{equation}
which is suppressed by the square root of the volume factor.
With $g v_N \leftrightarrow \Lambda_{\rm comp}$ and $N \sim \pi kR$, these two expressions are consistent.

\subsection{Higgs physics} \label{subsec:Higgs}

\renewcommand{\arraystretch}{1.3}
\begin{table}
\vspace{-0.3cm}
\begin{center}
\makeatletter
\def\@captype{table}
\makeatother
\begin{tabular}[t]{c|c|c}
& $SU(3)_C \times SU(2)_L \times U(1)_Y$ & $SU(3)_S$ \\
\hline
$f_u$ & $({\bf 1}, {\bf 2}, 1/2)$  & ${\bf 3}$  \\
$\tilde{f}_d$ & $({\bf 1}, {\bf 2}, -1/2)$  & ${\bf \bar{3}}$  \\
$f_0$ & $({\bf 1}, {\bf 1}, 0)$  & ${\bf 3}$  \\
$\tilde{f}_0$ & $({\bf 1}, {\bf 1}, 0)$  & ${\bf \bar{3}}$  \\
$f_p \, (p = 1,2,3)$ & $({\bf 1}, {\bf 1}, 0)$  & ${\bf 3}$  \\
$\tilde{f}_p \, (p = 1,2,3)$ & $({\bf 1}, {\bf 1}, 0)$  & ${\bf \bar{3}}$  \\
\end{tabular}
\end{center}
\vspace{2mm}
 \caption{A UV model of $\lambda$-SUSY.}
\label{Higgsmodel}
\vspace{0.5cm}
\end{table} 
\renewcommand{\arraystretch}{1.3}

In the 5D warped natural SUSY framework, the $125 \, \rm GeV$ Higgs mass is addressed by so-called $\lambda$-SUSY \cite{Barbieri:2006bg},
where we introduce a singlet field $S$ localized toward the IR brane
and form a superpotential coupling,
\begin{equation}
\begin{split}
\\[-2.5ex]
W_{\lambda} \,=\, \lambda S H_u H_d \, , \label{lambdasusy} \\[1.5ex]
\end{split}
\end{equation}
on the IR brane.
Since running of the 4D effective coupling constant $\lambda_{\rm eff}$ is cut off at the IR scale,
we can have a sizable $\lambda_{\rm eff}$ at the electroweak scale without encountering a Landau pole,
which enables us to lift up the Higgs mass easily.
On the other hand, deconstruction provides a UV theory beyond the IR scale, and hence
it is necessary to also present a UV completion of $\lambda$-SUSY.
As the existence proof, we comment on a model with a new supersymmetric QCD sector similar to the model presented in Ref.~\cite{Chang:2004db},
generating a composite singlet chiral superfield $S$ coupled to the Higgs fields as in \eqref{lambdasusy}.

Let us consider a supersymmetric $SU(3)_S$ gauge theory with $6$ vector-like flavors, $f$ and $\tilde{f}$.
The charge assignments are summarized in Table~\ref{Higgsmodel}. 
We introduce the following superpotential:
\begin{equation}
\begin{split}
W_{\rm Higgs} \,=\, \lambda_u H_u \tilde{f}_d f_0 + \lambda_d H_d f_u \tilde{f}_0 + m_f f_u \tilde{f}_d  \, , \\[1ex]
\end{split}
\end{equation}
where we only assume technical naturalness (e.g. we have ignored the mass terms of $f_0, \tilde{f}_0$ and $f_p, \tilde{f}_p$). 
The theory is in the middle of the conformal window
\cite{Intriligator:1995au}.
Even if the coupling constants $\lambda_u, \lambda_d$ are somewhat large,
their running does not hit a Landau pole because the strong gauge coupling of the new gauge theory
tends to make $\lambda_u$ and $\lambda_d$ smaller at high energies.
Integrating out $f_u, \tilde{f}_d$ below their mass scale $m_f$, the effective superpotential is generated as
\begin{equation}
\begin{split}
W_{\rm Higgs, \, eff} \,=\, - \frac{\lambda_u \lambda_d}{m_f} f_0 \tilde{f}_0 \, H_u H_d  \, . \\[1ex]
\end{split}
\end{equation}
The effective theory is an $SU(3)_S$ gauge theory with $4$ vector-like flavors,
$f_I = (f_0, \, f_p)$, $\tilde{f}_I = (\tilde{f}_0, \, \tilde{f}_p)$ ($I = 0, \cdots, 3$).
Below the dynamical scale $\Lambda_{\rm eff} \sim m_f$, this theory is described in terms of gauge invariant mesons, $M_{IJ} \sim f_I\tilde{f}_J$,
and baryons, $B^I \sim \epsilon^{IJKL} f_J f_K f_L$ and $\widetilde{B}^I \sim \epsilon^{IJKL} \tilde{f}_J \tilde{f}_K \tilde{f}_L$.
A component of the mesons, $M_{00} \sim f_0 \tilde{f}_0$, can be identified as the singlet chiral superfield $S$ in \eqref{lambdasusy}.
From NDA
\cite{Luty:1997fk}, we estimate the size of the $\lambda$ coupling as
\begin{equation}
\begin{split}
\lambda \,\sim\, \frac{\lambda_u \lambda_d}{4\pi} \frac{\Lambda_{\rm eff}}{m_f} \, . \\[1.5ex]
\end{split}
\end{equation}
When $\lambda$ is sizable, we are able to lift up the Higgs mass.
Further studies of this model and other possible UV models are left to a future work.

\section{Emergent supersymmetry} \label{sec:accidentalSUSY}

Warped gauge dynamics shows emergence of supersymmetry near the IR brane
even if SUSY is badly broken on the UV brane.
We first review the RG flow of the gauge theory coupled to a SCFT, the 4D dual to the 5D gauge theory, following Ref.~\cite{Sundrum:2009gv}.
Then, we present the RG flow of the gauge, gaugino and $D$-term couplings in a supersymmetric gauge moose theory
and demonstrate the phenomenon of emergent supersymmetry in the gauge moose theory.
Finally, the degree of fine-tuning is estimated in the deconstructed warped natural SUSY model.

\subsection{The gauge theory coupled to a SCFT} \label{subsec:CFT}

Let us consider a supersymmetric $SU(n)$ gauge theory coupled to a large-$N_{\rm CFT}$ superconformal field theory.
We assume that the SCFT is strongly-coupled and has the fewest possible relevant scalar operators
(The scalar operator like $|\phi|^2$ has a scaling dimension larger than $4$).
Then, the Lagrangian can be described by the following gauge ($g_A$), gaugino ($\tilde{g}$) and $D$-term ($g_D$) couplings:
\begin{equation}
\begin{split}
\\[-2.5ex]
\mathcal{L}_{J} \,\,=\,\, \mathcal{L}_{\rm SCFT} - \frac{1}{4} F_{\mu\nu}^a F_{\mu\nu}^a -i \bar{\lambda}^a \bar{\sigma}_\mu D_\mu \lambda^a
+ g_A A_\mu^a J_\mu^a \, +  \tilde{g} \, \lambda^a \Psi_J^a \, - \frac{1}{2} \, g_D^2 \, D_J^a D_J^a  \label{SCFTLagrangian}  \, , \\[1ex]
\end{split}
\end{equation}
where $\mathcal{L}_{\rm SCFT}$ is the Lagrangian of the SCFT and $J^a_\mu$, $\Psi_J^a$, $D_J^a$ are the associated CFT operators of a SUSY Yang-Mills.
The index $a$ denotes the gauge index.
If supersymmetry is unbroken, these couplings are same, $g_A = \tilde{g} = g_D$.
Instead, we consider the gauge theory coupled to matter chiral superfields whose scalars have large SUSY breaking masses.
Integrating out these scalars at their mass scale $m_0$, the effective theory only has the terms of the matter fermions,
\begin{equation}
\begin{split}
\\[-2.5ex]
\mathcal{L}_{\rm fermion} \,\,=\,\, -i \bar{\psi} \bar{\sigma}_\mu D_\mu \psi  \, , \\[1ex]
\end{split}
\end{equation}
in addition to the Lagrangian \eqref{SCFTLagrangian}.
Then, the RG flow of the effective theory does not respect supersymmetry
so that the gauge, gaugino and $D$-term couplings are split.

We now present the 1-loop renormalization group equations (RGEs) of the effective theory in the large-$N_{\rm CFT}$ expansion.
The running of the gauge coupling is
\begin{equation}
\begin{split}
\\[-2.5ex]
\frac{d }{d \ln \mu} \left( \frac{1}{g_A^2} \right) \,=\, - \, ( b_{\rm CFT} + b_{\, \rm fermion} ) \, , \label{gaugeRGEinverse} 
\end{split}
\end{equation}
or
\begin{equation}
\begin{split}
\frac{d g_A^2}{d \ln \mu} \,=\, (b_{\rm CFT} + b_{\, \rm fermion}) \, g_A^4 \, , \label{gaugeRGE} \\[1ex]
\end{split}
\end{equation}
where $b_{\rm CFT}$ is the beta function contribution from the SCFT sector
and $b_{\, \rm fermion}$ is the contribution from the fermion components of the matter chiral superfields.
We have ignored the subdominant contribution from the $SU(n)$ gauge fields, $b_{\, \rm SYM} = 3n/(8\pi^2)$.
In addition, the RGE of the gaugino coupling is given by
\begin{equation}
\begin{split}
\\[-2.5ex]
\frac{d \tilde{g}^2}{d \ln \mu} \,=\, b_{\rm CFT} \, \tilde{g}^4 \, . \label{gauginoRGE} \\[1ex]
\end{split}
\end{equation}
Note that there is no contribution from the matter fermions $b_{\, \rm fermion}$ in this equation at 1-loop.
The contribution from the CFT sector is the same as that of the gauge coupling, reflecting the fact that this sector is supersymmetric.
Finally, the RGE of the $D$-term coupling is
\begin{equation}
\begin{split}
\frac{d g_D^2}{d \ln \mu} \,=\, b_{\rm CFT} \, g_D^4 - \gamma_D (g_D^2 - g_A^2) \, . \label{D-termRGE} \\[1.5ex]
\end{split}
\end{equation}
The contribution from the CFT sector is the same as those of $g_A$ and $\tilde{g}$.
The important difference from the gauge and gaugino couplings is the second term proportional to $(g_D^2 - g_A^2)$.
If supersymmetry was preserved, this term would vanish.
The dimensionless coefficient $\gamma_D$ is $\mathcal{O}(1/ N_{\rm CFT})$ and the 5D calculation tells us that $\gamma_D \propto n/b_{\rm CFT}$.

The leading contribution in the RGEs of the SUSY Yang-Mills couplings
is the SCFT contribution $b_{\rm CFT} = \mathcal{O}(N_{\rm CFT} / 16\pi^2)$
and hence all these couplings are IR-free.
The $b_{\rm CFT}$ contribution dilutes the non-supersymmetric effects such as
the fermion contribution $b_{\, \rm fermion}$ in \eqref{gaugeRGE} and the term proportional to $(g_D^2 - g_A^2)$ in \eqref{D-termRGE}.
Then, the growth of splittings among the couplings is suppressed in the RG flow.

\subsection{The supersymmetric gauge moose} \label{subsec:CFT}

We now consider the RG flow of a supersymmetric $SU(n)^{N+1}$ gauge theory as shown in the moose diagram of Figure~\ref{fig:EffectiveMoose},
corresponding to the above gauge theory coupled to a SCFT.
At the UV end, the matter fermions live without their superpartners and supersymmetry is explicitly broken.

\subsubsection{The gauge coupling} \label{subsubsec:gauge}

Due to the hierarchical VEVs \eqref{vev2}, we can easily derive the running equation for the gauge coupling
by considering the RG flow and the gauge symmetry breaking step by step
\cite{Falkowski:2002cm}:
the gauge coupling of $G_{(j-1)}$ at the scale $v_{j-1}$
runs down to the scale $v_j$ and
then $G_{(j-1)} \times G_j$ is broken to the diagonal subgroup $G_{(j)}$.
The RGE of the gauge coupling is given by
\begin{equation}
\begin{split}
\frac{1}{g_{(j)}^2 \, (v_j)} \,&=\, \frac{1}{g_{(j-1)}^2 \, (v_{j-1})} + \frac{1}{g^2} + b_{\, \rm fermion} \ln \left( \frac{v_{j-1}}{v_j} \right) \\[2ex]
\,&=\, \frac{1}{g_{0}^2} + \frac{j}{g^2} + b_{\, \rm fermion} \ln \left( \frac{v}{v_j} \right) \, , \label{gaugeRGEmoose} \\[2ex]
\end{split}
\end{equation}
where $g_0$ is the gauge coupling of the UV end site
and $b_{\, \rm fermion}$ is the contribution from the fermion components of the matter chiral superfields as before.
We have ignored the subleading contribution from the gauge fields $b_{\rm SYM}$ and the contribution from the link fields.

In section~\ref{sec:Moose/CFT}, we have discussed the correspondence between $b_{\rm CFT}$ and the gauge coupling of a moose theory $g$,
which is given by \eqref{bcft/gauge}.
The above RGE of the gauge coupling can be rewritten as
\begin{equation}
\begin{split}
\\[-2.5ex]
\frac{\Delta_j {1}/{g_{(j)}^2}}{\Delta_j \ln v_j}  \, \equiv \, \frac{ {1}/{g_{(j-1)}^2 \, (v_{j-1})} - {1}/{g_{(j)}^2 \, (v_j)}}{ \ln{v_{j-1}} - \ln{v_{j}} }
\,=\,  - \frac{1}{g^2} \left( \frac{k}{gv} \right)^{-1} - b_{\, \rm fermion} \, . \\[1ex]
\end{split}
\end{equation}
Comparing with \eqref{gaugeRGEinverse}, we easily see the correspondence.
As in the CFT calculation, the leading contribution is the first supersymmetric term.
This contribution dilutes the non-supersymmetric effect from the second term.

\subsubsection{The gaugino coupling} \label{subsubsec:gaugino}

The running of the gaugino coupling can be obtained as well
by considering the RG flow and the gauge symmetry breaking step by step.
For the gaugino coupling, there is no contribution from the matter fermions $b_{\, \rm fermion}$ at 1-loop.
The result is shown as
\begin{equation}
\begin{split}
\\[-2.5ex]
\frac{1}{\tilde{g}_{(j)}^2 \, (v_j)} \,&=\, \frac{1}{\tilde{g}_{(j-1)}^2 \, (v_{j-1})} + \frac{1}{g^2}  \\[2ex]
\,&=\, \frac{1}{\tilde{g}_{0}^2} + \frac{j}{g^2} \, , \label{gauginoRGEmoose} \\[1ex]
\end{split}
\end{equation}
where $\tilde{g}_0$ is the gaugino coupling at the UV end site, which deviates from the gauge coupling $g_0$ in general because SUSY is broken at this site.
Note that the contribution $1/g^2$ is equal to the one appeared in the RGE of the gauge coupling \eqref{gaugeRGEmoose},
consistent with the fact that the CFT sector is supersymmetric.
From \eqref{gaugeRGEmoose} and \eqref{gauginoRGEmoose}, we obtain
\begin{equation}
\begin{split}
\\[-2.5ex]
\frac{\Delta \tilde{g}_{(j)}^2}{g_{(j)}^2} \,\equiv \, \frac{\tilde{g}_{(j)}^2 - {g}_{(j)}^2}{g_{(j)}^2}
\,\simeq\, \frac{g_{(j)}^2}{g_0^2} \frac{\Delta \tilde{g}_{0}^2}{g_{0}^2} + b_{\, \rm fermion} \, g_{(j)}^2 \ln \left( \frac{v}{v_j} \right) \, . \label{gauginosplitting} 
\end{split}
\end{equation}
Since the gaugino coupling is IR-free due to the contribution $1/g^2$, we have $g_{(j)}^2 < g_0^2$ for $j > 0$ so that
the initial splitting $\Delta \tilde{g}_{0}^2 / g_{0}^2$ is diluted at low energies. This is an important aspect of emergent supersymmetry.

\subsubsection{The D-term coupling} \label{subsubsec:gaugino}

To trace the running of the $D$-term coupling, we introduce
a vector-like pair of spectator chiral superfields $\Phi_j, \widetilde{\Phi}_j$ ($j = 0, \cdots, N$), which has the (anti-)fundamental representation under each gauge group of the moose,
and define the (squared of) $D$-term coupling as the coupling constant of the scalar quartic term
(The spectator fields are assumed to be integrated out together with the gauge fields).
When supersymmetry is preserved, this is identical with the gauge coupling of the vector superfield.
As with the gauge and gaugino couplings presented above, the tree-level contribution to the $D$-term coupling
at the scale $v_j$ is given by
\begin{equation}
\begin{split}
\frac{1}{{g}_{D(j), \, \rm tr}^2} \,=\, \frac{1}{{g}_{D(j-1)}^2} + \frac{1}{g^2} \, , \label{D-termtree} \\[1.5ex]
\end{split}
\end{equation}
and the supersymmetric correction $1/g^2$ corresponds to the SCFT contribution.
On the other hand, there is a $(g_D^2 - g_A^2)$ correction in the RGE of the CFT picture \eqref{D-termRGE}.
The dimensionless coefficient $\gamma_D$ is proportional to $1/b_{\rm CFT}$.
What is the counterpart of this correction in the gauge moose theory?
Since the correspondence in section~\ref{sec:Moose/CFT} says $b_{\rm CFT} \sim 1/g^2$,
the expected term in the RG flow of the moose theory is proportional to $g^2$.
In fact, we find this correction as we prove next.

Since $G_{(j-1)} \times G_j$ is broken to the diagonal subgroup $G_{(j)}$ at the scale $v_j$,
a linear combination of the vector superfields, $V^{(j-1)}$ and $V^j$, becomes massive.
In the supersymmetric limit, substituting the link field VEVs,
the Lagrangian of $V^{(j-1)}$, $V^j$ and the spectator chiral superfields $\Phi_j, \widetilde{\Phi}_j$ can be expanded as
\begin{equation}
\begin{split}
\\[-2.5ex]
&\int d^4 \theta \left[ \, \frac{v_j^2}{2} \left( {\rm Tr} \, e^{-g V^j} e^{\, g_{(j-1)} V^{(j-1)}}
+ {\rm Tr} \, e^{-g_{(j-1)} V^{(j-1)}}  e^{\, g V^j} \right)  + K_{\rm sp} \, \right]  , \\[1.5ex]
=\, &\int d^4 \theta \left[ \, \frac{v_j^2}{2} \left\{ \, 2 n + {\rm Tr} \left( \, g^2 (V^{j})^2
+ g_{(j-1)}^2 (V^{(j-1)})^2 - 2 g \, g_{(j-1)} V^j V^{(j-1)} \, \right) + \cdots \right\}  + K_{\rm sp} \, \right] , \label{expandedlag} \\[1ex]
\end{split}
\end{equation}
where $K_{\rm sp} ( \Phi_j, \widetilde{\Phi}_j )$ denotes the Kahler potential of the spectator chiral superfields.
We can define the mass eigenstates of the vector superfields as
\begin{equation}
\begin{split}
\\[-2ex]
V^{(j)} \, \equiv \, \frac{g_{(j-1)} V^j + g \, V^{(j-1)}}{\sqrt{g_{(j-1)}^2 + g^2}} \, ,
\qquad \widetilde{V}^{(j)} \, \equiv \, \frac{g \, V^j - g_{(j-1)} V^{(j-1)}}{\sqrt{g_{(j-1)}^2 + g^2}}  \, . \label{eigenstates} 
\end{split}
\end{equation}
Here, $V^{(j)}$ and $\widetilde{V}^{(j)}$ are the massless and massive eigenstates respectively.
Substituting \eqref{eigenstates} into \eqref{expandedlag}
and expanding the spectator Kahler potential $K_{\rm sp}$,
the Lagrangian is rewritten as
\begin{equation}
\begin{split}
&\int d^4 \theta \,  \frac{v_j^2}{2} \left( 2 n + \frac{1}{2} (g_{(j-1)}^2 + g^2) \, \widetilde{V}^{(j) a} \, \widetilde{V}^{(j) a} + \cdots \right)  \\[2ex]
&+ \int d^4 \theta \, \biggl( \Phi_j^{\dagger} \Phi_j - \frac{g^2}{ \sqrt{g_{(j-1)}^2 + g^2} } \, \Phi^{\dagger}_j \, \widetilde{V}^{(j) a} T^a \, \Phi_j + \cdots \biggr) \, . \\[1ex]
\end{split}
\end{equation}
We have used $\tr \,{(T^a T^b)} = \frac{1}{2} \delta^{ab}$.
We now integrate out the massive vector superfield $\widetilde{V}^{(j)}$.
Using the equation of motion of $\widetilde{V}^{(j)}$ obtained from the above Lagrangian,
the effective Kahler potential of the spectator chiral superfield $\Phi_j$ is given by
\begin{equation}
\begin{split}
\\[-2.5ex]
K_{\rm eff} \, = \, \Phi_j^{\dagger} \Phi_j  \, - \, \frac{1}{2}  \frac{2 g^4}{(g_{(j-1)}^2 + g^2)^2 v_j^2} \, \bigl| \Phi_j^{\dagger} \, T^a \Phi_j \bigr|^2 \, , \label{specD} \\[1ex]
\end{split}
\end{equation}
where we omit writing the dependence on the massless vector superfield $V^{(j)}$ explicitly.
Note that $g_{(j-1)}$ in this expression is the gauge coupling of $G_{(j-1)}$ at the scale $v_j$.

Splittings between the gauge, gaugino and $D$-term couplings lead to
a SUSY breaking radiative correction in the massive vector multiplet.
As in the case of lifting up the Higgs boson mass by non-decoupling $D$-terms \cite{Batra:2003nj},
this leads to generation of the quartic coupling of the spectator scalar fields.
To include the SUSY breaking effect,
we make the replacement, $v_j^2 \rightarrow v_j^2 (1 - m_{\rm SUSY}^2 \theta^4)$, in the above calculation.
Then, from \eqref{D-termtree} and \eqref{specD},
we obtain the $D$-term coupling at the scale $v_j$,
\begin{equation}
\begin{split}
\\[-2.5ex]
g_{D (j)}^2 \, = \, g_{D (j), \, \rm tr}^2 + \frac{2 g_{(j)}^4 }{g_{(j-1)}^4  } \frac{m_{\rm SUSY}^2}{v_j^2} \, . \label{gd_msusy} \\[1ex]
\end{split}
\end{equation}
The mass-squared parameter $m_{\rm SUSY}^2$ encapsulates the SUSY breaking mass of the link scalar field which becomes a part of the massive vector multiplet.
Since the scalar is adjoint under the diagonal subgroup $G_{(j)}$,
we can estimate $m_{\rm SUSY}^2$ as
\begin{equation}
\begin{split}
\\[-2.5ex]
m_{\rm SUSY}^2
\, \sim \, \frac{n}{16\pi^2} \, \Delta g_{D (j)}^2  \,  k_{j}^2 \, , \label{msusy} \\[1ex]
\end{split}
\end{equation}
where $\Delta g_{D (j)}^2 \equiv g_{D (j)}^2 - g_{(j)}^2$ is the splitting between the gauge coupling and the $D$-term coupling.
We have ignored the subleading effect from the difference between the gauge and gaugino couplings
because this splitting is of order $g_{(j)}^4$ by \eqref{gauginosplitting}.
The factor $n$ comes from the quadratic Casimir of the adjoint representation under $SU(n)$.
Since the warped background suggests that the supersymmetric Lagrangian of the vector-like pair of link chiral superfields $\Sigma_j, \widetilde{\Sigma}_j$ is defined around the scale of their VEVs,
the quadratic divergence is cut off at the scale $g v_j \sim k_{j} \equiv k e^{-kj/gv}$.
Inserting \eqref{msusy} into the $D$-term coupling \eqref{gd_msusy}, we finally obtain
\begin{equation}
\begin{split}
\frac{1}{g_{D (j)}^2} \, \simeq \, \frac{1}{g_{D (j-1)}^2} + \frac{1}{g^2}
- \frac{n}{8\pi^2} \frac{g^2 \Delta g_{D (j)}^2 }{g_{(j-1)}^4  }  \left( \frac{k}{gv} \right)^2  \, . \label{hard} \\[1.5ex]
\end{split}
\end{equation}
In this expression, we approximate $\Delta g_{D (j)}^2$ by the splitting between the gauge coupling and the $D$-term coupling at tree-level,
$\Delta g_{D (j)}^2 \simeq g_{D (j) , \, \rm tr}^2 - g_{(j)}^2$.
Comparing this equation with the CFT calculation \eqref{D-termRGE}, we easily see ``the coefficient $\gamma_D$" of the moose theory,
\begin{equation}
\begin{split}
\\[-2.5ex]
\gamma_{D, \, \rm dec} \, \equiv \, \frac{n}{8\pi^2} \, g^2 \left( \frac{k}{gv} \right)  
 \, \leftrightarrow \,  \frac{n}{8 \pi^2 \, b_{\rm CFT}}   \, , \\[1.5ex]
\end{split}
\end{equation}
where we have used the correspondence \eqref{bcft/gauge}.
As expected, this is proportional to $1/b_{\rm CFT}$.\footnote{
The apparent discrepancy of the numerical factor with Eq.(23) in \cite{Sundrum:2009gv} comes from
naive dimensional analysis in \eqref{msusy}.}
For a $U(1)$ gauge group, the link scalar is singlet under the diagonal subgroup $G_{(j)}$ and hence $\gamma_{D, \, \rm dec}$ vanishes, consistent with the 5D calculation
\cite{Sundrum:2009gv}.

\subsection{Fine tuning} \label{subsec:finetuning}

Let us analyze the degree of fine-tuning in the warped natural SUSY framework presented in the previous section.
The Higgs fields live at the IR end of the gauge moose.
We consider the case where all the third generation quark and lepton multiplets are localized toward the IR brane
and take the simplification that they couple to only the gauge group $G_N$ in the deconstructed model.
We expect that there is no significant effect due to putting them in the bulk of the gauge moose.
The superpartners of light quarks and leptons have large SUSY breaking masses,
\begin{equation}
\begin{split}
m_0 \, \sim \, \frac{F^2}{M^2} \, , \\[1ex]
\end{split}
\end{equation}
where $F$ denotes the $F$-term of a SUSY breaking spurion and $M$ is some mediation scale.
Splittings between the gauge, gaugino and $D$-term couplings at the scale $M$ is estimated to be $\sim F^2/M^4$.
If we assume $F < M^2$, the splittings are small and further suppressed by the effect discussed below \eqref{gauginosplitting}
and the similar effect expected in the RGE of the $D$-term coupling.
Then, as in Ref.~\cite{Sundrum:2009gv}, we take $\Delta \tilde{g}_{0}^2 = \Delta g_{D 0}^2 = 0$ at the UV end of the effective moose theory below the scale $m_0$.

While the RGEs of the gauge, gaugino and $D$-term couplings are non-supersymmetric due to the absence of the scalar superpartners of light quarks and leptons,
the moose of gauge groups dilutes the SUSY breaking effect.
The remaining hard SUSY breaking is given by
splittings between the gauge coupling \eqref{gaugeRGEmoose}, the gaugino coupling \eqref{gauginoRGEmoose} and the $D$-term coupling \eqref{hard}
of the unbroken diagonal subgroup.
They lead to a quadratic divergence in the mass of the IR-end localized scalar, the fundamental representation of the $SU(n)$ gauge group, which is cut off at the IR scale,
\begin{equation}
\begin{split}
\\[-2.5ex]
\Delta m^2 \,\sim\, \frac{n^2-1}{2n} \frac{\Delta g_{D(N)}^2}{16\pi^2} \, \Lambda_{\rm comp}^2 \, , \label{quadraticdivergence} \\[1.5ex]
\end{split}
\end{equation}
where $\Delta g_{D (N)}^2$ is the splitting between the gauge coupling and the $D$-term coupling of the unbroken diagonal subgroup.
Although the Yukawa couplings of light quarks and leptons are non-supersymmetric, these couplings are small and their effects on fine-tuning are negligible.

\begin{figure}[t]
 \begin{center}
  \subfigure[\label{sfig:gauge_2}]{\includegraphics[clip, width=6cm]{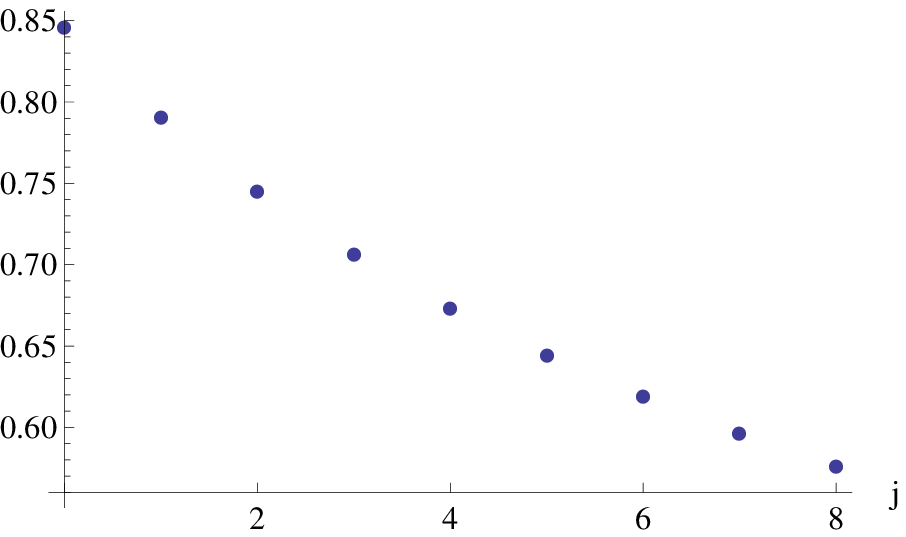}}\hspace{15mm}
    \subfigure[\label{sfig:gauge_3}]{{\includegraphics[clip, width=6cm]{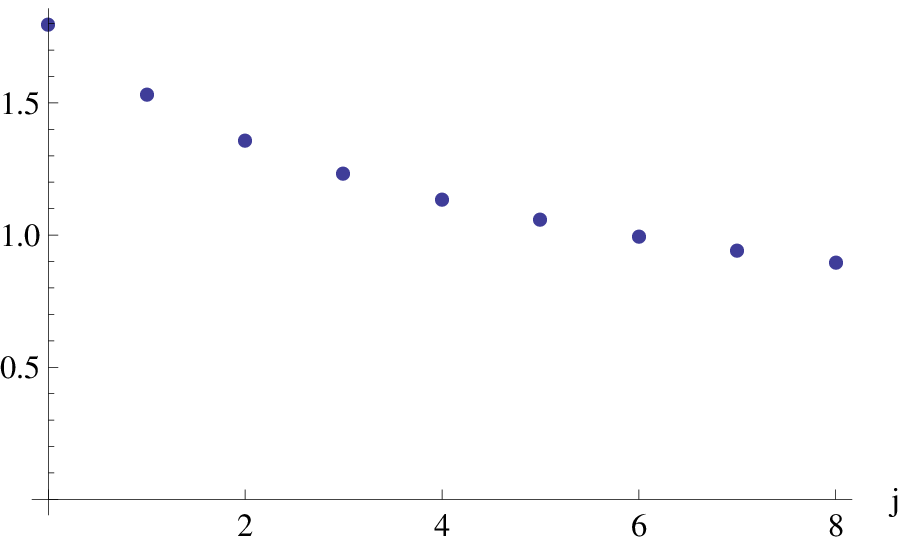}}}
    \caption{The running of the gauge couplings $g_{(j)}$ for \subref{sfig:gauge_2}~the $SU(2)_L$ gauge group
and \subref{sfig:gauge_3}~the $SU(3)_C$ gauge group. 
The UV cutoff scale is $m_0 = 10^4 \, \rm TeV$ and the IR scale is $\Lambda_{\rm comp} = 10 \, \rm TeV$.
We take $N=8$ and $b_{\rm CFT}^{(n=2)} = 0.2$, $b_{\rm CFT}^{(n=3)} = 0.1$.
\label{fig:gaugeRGE}}
\end{center}
\end{figure}

\begin{figure}[t]
 \begin{center}
  \subfigure[\label{sfig:n_2}]{\includegraphics[clip, width=6cm]{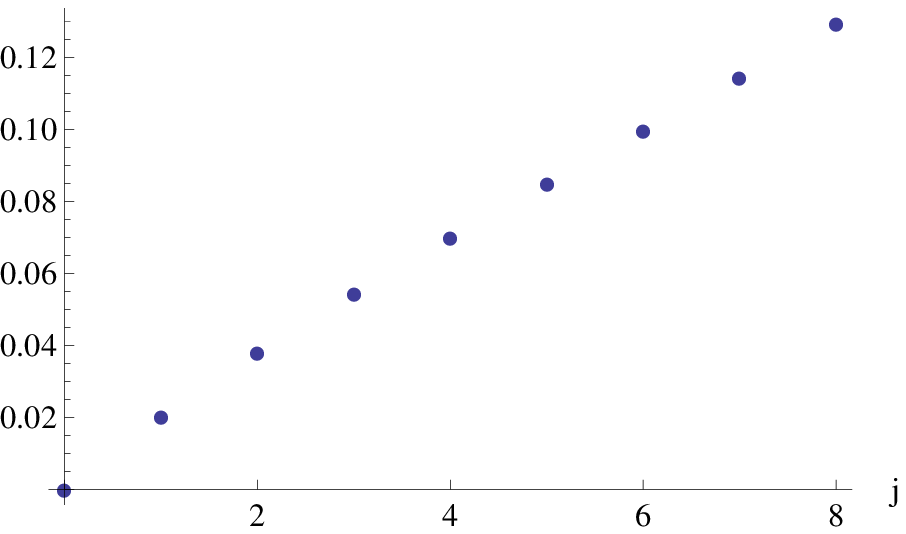}}\hspace{15mm}
    \subfigure[\label{sfig:n_3}]{{\includegraphics[clip, width=6cm]{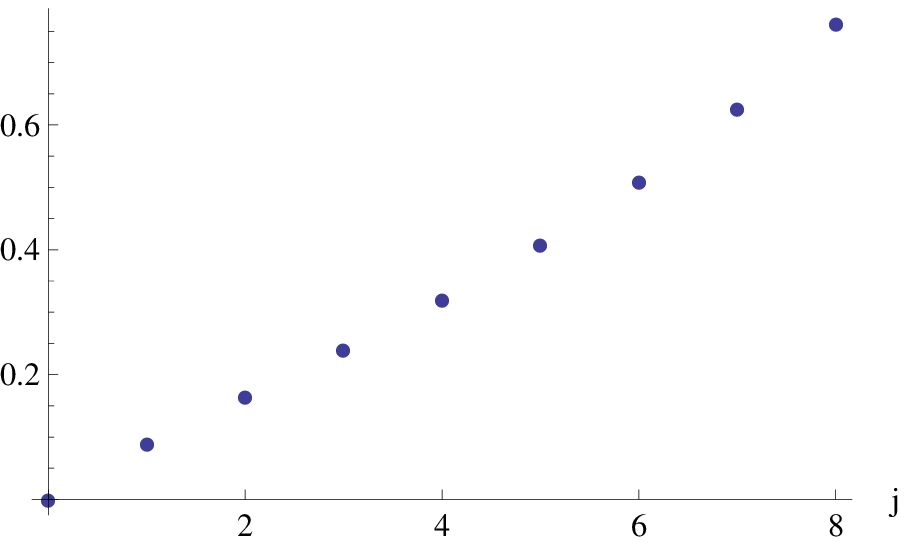}}}
    \caption{The growth of $\Delta g_{D (j)}^2 / g_{(j)}^2$ for \subref{sfig:n_2}~the $SU(2)_L$ gauge group
and \subref{sfig:n_3}~the $SU(3)_C$ gauge group. We take $\Delta \tilde{g}_{0}^2 = \Delta g_{D 0}^2 = 0$. 
\label{fig:numerical}}
\end{center}
\end{figure}

We now estimate the size of the quadratic divergence \eqref{quadraticdivergence} numerically.
We assume that the UV cutoff scale, corresponding to the mass scale of the heavy scalars, is $m_0 = 10^4 \, \rm TeV$
and the IR scale is $\Lambda_{\rm comp} = 10 \, \rm TeV$.
We take the number of the gauge sites as $N=8$ for every gauge group.
Then, according to the correspondence \eqref{bcft/gauge}, the gauge coupling of the moose $g$ is determined by
the beta function coefficient from the CFT sector $b_{\rm CFT}$.
Suppose that $b_{\rm CFT}^{(n=2)} = 0.2$ ($g^{(n=2)} \simeq 2$) for the $SU(2)_L$ group
and $b_{\rm CFT}^{(n=3)} = 0.1$ ($g^{(n=3)} \simeq 3$) for the $SU(3)_C$ gauge group.
The gauge couplings of the unbroken diagonal subgroups at the IR scale are taken as $g_{(N)}^{(n=2)} \simeq 0.6$ and $g_{(N)}^{(n=3)} \simeq 1$.
Then, the couplings at the UV cutoff scale $m_0$ are given by $g_0^{(n=2)} \simeq 1$ and $g_0^{(n=3)} \simeq 2$ without Landau poles.
The beta function coefficient from two light generations of quarks and leptons are given by $b_{\, \rm fermion}^{(n=2)} = b_{\, \rm fermion}^{(n=3)} = 1/3\pi^2$.
The running of the gauge couplings $g_{(j)}$ for \subref{sfig:gauge_2}~the $SU(2)_L$ gauge group
and \subref{sfig:gauge_3}~the $SU(3)_C$ gauge group is plotted in Figure~\ref{fig:gaugeRGE}.
This corresponds to the IR-free behavior of the gauge couplings in the CFT case.
Figure~\ref{fig:numerical} shows the growth of $\Delta g_{D (j)}^2 / g_{(j)}^2$ for \subref{sfig:n_2}~the $SU(2)_L$ gauge group
and \subref{sfig:n_3}~the $SU(3)_C$ gauge group.
Although they start from zero and increase as $j$ becomes large, their growth is suppressed by the supersymmetric contribution $1/g^2$ and $\Delta g_{D (N)}^2 / g_{(N)}^2 < 1$
for both gauge groups.
Then, by using \eqref{quadraticdivergence}, the corrections to the soft masses of the stop and the Higgs are estimated as
\begin{equation}
\begin{split}
\\[-2.5ex]
\Delta m_{ \tilde{t}}^2 \, \simeq \, (720 \, {\rm GeV})^2 \, , \qquad  \Delta m_H^2 \simeq (140 \, {\rm GeV})^2 \, . \label{tuning} \\[1ex]
\end{split}
\end{equation}
For reference, if there was no dilution, $\Delta g_{D (N)}^2 / g_{(N)}^2 \sim 1$, the corrections would be given by
$\Delta m_{ \tilde{t}}^2 \, \simeq \, (920 \, {\rm GeV})^2$ for the stop mass and $\Delta m_H^2 \simeq (410 \, {\rm GeV})^2$ for the Higgs soft mass.
The stop mass in \eqref{tuning} contributes to the Higgs soft mass through the top/stop loop, but the required tuning is only around $10 \%$.\footnote{
In addition, there is a negative contribution to the scalar mass from a two-loop quadratic divergence induced by light quarks and leptons below the IR scale $\Lambda_{\rm comp}$
\cite{Heidenreich:2014jpa}
and hence further relaxation of tuning is possible.}

\section{Conclusions} \label{sec:conclusions}

We have studied a gauge theory in a 5D warped space via deconstruction
and presented the correspondence between a moose of gauge groups and a CFT,
including the relation of the gauge coupling of the moose with the beta function coefficient from the CFT.
Furthermore, a supersymmetric extension of deconstruction has been proposed.
In the deconstructed version of warped natural SUSY, 
a supersymmetry breaking source is located at the UV end of the moose.
Since the third generation quark and lepton multiplets are localized at the IR end,
they do not couple to the SUSY breaking source.
The superpartners of light quarks and leptons are decoupled.
This may lead to the $U(1)$ $D$-term problem, which can be understood in terms of the moose theory.
The solution has been provided by a left-right symmetric gauge moose.
The model predicts light $W'$ and $Z'$ gauge fields tested at the LHC.
The proposed dictionary of ``Moose/CFT correspondence'' helps our understanding of the RG flow and realization of emergent supersymmetry.
Although the superpartners of light quarks and leptons are decoupled,
the quadratic divergence in the mass of the scalar living at the IR end is suppressed due to supersymmetry in the bulk of the moose,
which stabilizes the little hierarchy between the IR scale $\Lambda_{\rm comp}$ and the electroweak scale
with light stops, Higgsinos and gauginos.

Possible future directions are as follows.
In this paper, we have assumed a coarse lattice so that the VEVs of the link scalars in the deconstructed theory are hierarchical.
On the other hand, the fine lattice limit has been discussed by several authors.
Ref.~\cite{deBlas:2006fz} investigated the gauge boson mass spectrum
and Ref.~\cite{Katz:2004qa} presented the RG flow of the gauge coupling in this limit.
We can consider a supersymmetric extension of fine-lattice deconstruction
and explore the RG flow.
It would be interesting to demonstrate emergence of supersymmetry in this limit.

In addition, we have not presented an explicit large-$N_c$ theory corresponding to
the RS model or its supersymmetric extension.
This might be provided from supersymmetric QCD (SQCD) or its SUSY breaking deformation.
The Seiberg duality has been established in SQCD.
While the hidden local symmetry is only a phenomenological model in real QCD,
the author of Ref.~\cite{Komargodski:2010mc} proposed that
this symmetry can be understood as the magnetic gauge symmetry in the Seiberg dual theory.
The link fields are identified as the dual quarks.
This idea was applied to real QCD with a SUSY breaking deformation in Ref.~\cite{Kitano:2011zk}.
Then, it might be possible to construct the correspondence between a supersymmetric warped model and a large-$N_c$ SQCD as in Ref.~\cite{Son:2003et}.
A SUSY breaking deformation of this AdS/SQCD relation could provide a new insight into the AdS/QCD.

Furthermore, we have assumed scalar, fermion or vector fields in the discussion of the relation between a moose theory and a CFT.
It might be possible to include gravity in this relation.
Discretization of gravity in a warped space was presented in \cite{Randall:2005me}.
A moose theory of gravity could provide a further connection among the two 4D theories.

\section*{Acknowledgements}
The author would like to thank Ben Heidenreich and Matthew Reece for valuable discussions and comments on the manuscript.
He is also grateful to Marat Freytsis, Masaki Murata and Ryosuke Sato for helpful conversations.
He is supported by JSPS Fellowships for Young Scientists.

\appendix
\renewcommand{\theequation}{A.\arabic{equation}}

\section*{Appendix: Deconstruction of bulk matter fields} \label{appendix}

In this appendix, we review deconstruction of bulk matter fields
\cite{Abe:2002rj}.
We consider a 5D bulk fermion coupled to a $U(1)$ gauge field.
This is composed of two Weyl spinors $\psi$, $\psi^c$ which have charges $1$, $-1$ respectively.
As in the case of the gauge field, we first present the continuum theory and latticize it.
The continuum action contains
\begin{equation}\label{fermioncontinuum}
\begin{split}
\\[-2.5ex]
S_{5, \, \rm fermion} \,\supset\, - \int d^4 x \int dy \, e^{-ky} \left( \psi^c \partial_5 \psi + \left( c - \frac{1}{2} \right) k \, \psi^c \psi + {\rm h.c.} \right), \\[1.5ex]
\end{split}
\end{equation}
where we have omitted the kinetic terms with the 4D derivative $\partial_\mu$ (and the gauge interactions), which are unimportant for the following discussion.
The (dimensionless) bulk mass parameter $c$ controls the wavefunction profile of the zero mode (See Ref.~\cite{Gherghetta:2010cj}).
We assume the following boundary condition on the UV brane:
\begin{equation}
\begin{split}
\\[-2.5ex]
\partial_5 \psi (0) \,=\, \psi^c (0) \,=\, 0 \, . \\[1.5ex]
\end{split}
\end{equation}
The latticized action is then given by
\begin{equation}\label{fermionlatticized}
\begin{split}
\\[-2.5ex]
- \int d^4 x \, a \, \sum_{j=1}^N e^{-k y_j} \left( \psi^c (y_j) \frac{\psi (y_j) -\psi (y_{j-1})}{a}
+ \left( c - \frac{1}{2} \right) k \, \psi^c (y_j) \psi (y_j) + {\rm h.c.} \right). \\[1.5ex]
\end{split}
\end{equation}
We now present a deconstructed model of this 5D theory.
A 5D bulk fermion is described in terms of spinors $\psi_j$, $\psi^c_j$ at every site of the moose in Figure~\ref{fig:moose}.
They have charges $1$, $-1$ under the $U(1)_j$ subgroup.
The action is presented as
\begin{equation}\label{fermiondeconstructed}
\begin{split}
\\[-2.5ex]
S_{4, \, \rm fermion} \, \supset \, \int d^4 x \, \sum_{j=1}^N \, \left[ \sqrt{2} g \, \psi_{j-1} \Sigma_j \psi^c_j
- \left\{ \frac{k}{gv} \left( c - \frac{1}{2} \right) +1  \right\} g v_j \, \psi^c_j \psi_j + {\rm h.c.} \right], \\[1.5ex]
\end{split}
\end{equation}
where $\Sigma_j$ are the link fields shown in section~\ref{sec:deconstruction}.
The correspondence between the spinors of the (latticized) 5D theory and those of the deconstructed theory is given by
\begin{equation}
\begin{split}
\\[-2.5ex]
\psi (y_j) \,\leftrightarrow\, (gv)^{1/2} \, \psi_j \, , \qquad \psi^c (y_j) \,\leftrightarrow\, (gv)^{1/2} \, \psi^c_j \, . \\[1.5ex]
\end{split}
\end{equation}
The action contains the fermion mass terms,
\begin{equation}
\begin{split}
\\[-2.5ex]
\mathcal{L}_{4, \, \rm fermion} &\, \supset \, \sum_{j=1}^N  \left( \, g v_j \, \psi_{j-1} \psi^c_j -   \xi^{-1} g v_j \, \psi^c_j \psi_j + {\rm h.c.} \right), \\[1ex]
\end{split}
\end{equation}
where we have defined
\begin{equation}
\begin{split}
\\[-2.5ex]
\xi^{-1} \, = \, \frac{k}{gv} \left( c - \frac{1}{2} \right) +1 \, . \\[1ex]
\end{split}
\end{equation}
As in the case of the gauge field, we can obtain the mass spectrum by diagonalizing the mass matrix of two adjacent sites step by step.
When the Weyl spinor $\psi$ ($\psi^c$) satisfies the Neumann (Dirichlet) condition on the IR brane,
we have one massless fermion,
\begin{equation}
\begin{split}
\\[-2.5ex]
\psi_{(N)} \, = \, \frac{1}{\sqrt{\sum_{j=0}^N \xi^{2j}}} \left( \, \sum_{j=0}^N  \xi^j \psi_j \right),  \\[1.5ex]
\end{split}
\end{equation}
which corresponds to the zero mode in the KK decomposition of the 5D fermion.
Finally, we present a supersymmetric extension of the above deconstructed model \eqref{fermiondeconstructed}.
The action contains the term,
\begin{equation}
\begin{split}
\\[-2.5ex]
-\int d^2 \theta \, \sum_{j=1}^N  \left( \sqrt{2} g \, \Psi_{j-1} \Sigma_j \Psi^c_j - \xi^{-1} g v_j \Psi_j^c \Psi_j \right) + {\rm h.c.} \, , \\[1ex]
\end{split}
\end{equation}
where $\Psi_j$, $\Psi_j^c$ are chiral superfields whose fermion components are spinors $\psi_j$, $\psi^c_j$.

\end{document}